\pdfoutput=1
\documentclass[11pt,twoside,a4paper,cmspaper,final,collab]{cms-tdr}
\def\svnVersion{89bf4cd}\def\svnDate{2024/11/06}\def\cmsCernNoTag{CERN-EP-2024-054}\def\cmsCernDate{\today}\def\cmsMessage{Published in Physical Review Letters as \href{http://dx.doi.org/10.1103/PhysRevLett.133.191902}{\doi{10.1103/PhysRevLett.133.191902}.}}
\begin{document}\cmsNoteHeader{EXO-23-002}

\newcommand{\PS}{\ensuremath{\text{S}}\xspace}
\newcommand{\mS}{\ensuremath{m_{\PS}}\xspace}
\newcommand{\Pphi}{\ensuremath{\phi}\xspace}
\newcommand{\mPhi}{\ensuremath{m_{\Pphi}}\xspace}
\DeclareRobustCommand{\PAprime}{{\HepParticle{A}{}{\prime}}\Xspace}
\newcommand{\mAprime}{\ensuremath{m_{\PAprime}}\xspace}
\newcommand{\BR}{\ensuremath{\mathcal{B}}\xspace}
\newcommand{\dark}{\ensuremath{\text{D}}\xspace}
\newcommand{\PqD}{\ensuremath{\text{q}_{\dark}}\xspace}
\newcommand{\mqD}{\ensuremath{m_{\PqD}}\xspace}
\newcommand{\lambdaD}{\ensuremath{\Lambda_{\dark}}\xspace}
\newcommand{\TD}{\ensuremath{T_{\dark}}\xspace}
\newcommand{\pipi}{\ensuremath{\PGpp\PGpm}\xspace}
\newcommand{\NAa}{\ensuremath{N_{\text{A}}}\xspace}
\newcommand{\NB}{\ensuremath{N_{\text{B}}}\xspace}
\newcommand{\NC}{\ensuremath{N_{\text{C}}}\xspace}
\newcommand{\NCi}{\ensuremath{\NC^{i}}\xspace}
\newcommand{\ND}{\ensuremath{N_{\text{D}}}\xspace}
\newcommand{\NE}{\ensuremath{N_{\text{E}}}\xspace}
\newcommand{\NF}{\ensuremath{N_{\text{F}}}\xspace}
\newcommand{\NFi}{\ensuremath{\NF^{i}}\xspace}
\newcommand{\NGg}{\ensuremath{N_{\text{G}}}\xspace}
\newcommand{\NH}{\ensuremath{N_{\text{H}}}\xspace}
\newcommand{\NSR}{\ensuremath{N_{\text{SR}}}\xspace}
\newcommand{\NSRi}{\ensuremath{\NSR^{i}}\xspace}
\newcommand{\abseta}{\ensuremath{\abs{\eta}}\xspace}
\newcommand{\lamtwo}{\ensuremath{\lambda_{2}}\xspace}
\newcommand{\lamthr}{\ensuremath{\lambda_{3}}\xspace}
\newcommand{\nSUEPConstituents}{\ensuremath{n_{\text{constituent}}^{\text{SUEP}}}\xspace}
\newcommand{\boostedSphericity}{\ensuremath{S^{\text{SUEP}}_{\text{boosted}}}\xspace}

\newlength\figWidth
\ifthenelse{\boolean{cms@external}}{\setlength{\figWidth}{0.99\linewidth}}{\setlength{\figWidth}{0.5\textwidth}}

\cmsNoteHeader{EXO-23-002}
\title{Search for soft unclustered energy patterns in proton-proton collisions at \texorpdfstring{13\TeV}{13 TeV}}

\date{\today}

\abstract{
The first search for soft unclustered energy patterns (SUEPs) is performed using an integrated luminosity of 138\fbinv of proton-proton collision data at $\sqrt{s}=13\TeV$, collected in 2016--2018 by the CMS detector at the LHC. Such SUEPs are predicted by Hidden Valley models with a new, confining force with a large 't Hooft coupling. In events with boosted topologies, selected by high-threshold hadronic triggers, the multiplicity and sphericity of clustered tracks are used to reject the background from standard model quantum chromodynamics. With no observed excess of events over the standard model expectation, limits are set on the cross section for production via gluon fusion of a scalar mediator with SUEP-like decays.
}

\hypersetup{%
pdfauthor={CMS Collaboration},%
pdftitle={Search for soft unclustered energy patterns in proton-proton collisions at 13 TeV},%
pdfsubject={CMS},%
pdfkeywords={CMS, dark QCD, dark matter, hidden valley, exotica, SUEP}}

\maketitle 

Recent experimental searches have not found evidence for beyond the standard model (BSM) physics.
Therefore, new models with unconventional experimental signatures that may have evaded previous constraints are being examined more closely.
In particular, Hidden Valley (HV) models~\cite{Cvetic:2002qa,Chacko:2005pe,Strassler:2006im,Hur:2007uz,Abel:2008ai,Bai:2013xga} predict a wide array of possible signals and have the potential to address unanswered questions in the standard model (SM), including the nature of dark matter~\cite{Bai:2013xga} and the difference between the vacuum expectation value of the Higgs field and the Planck scale, known as the hierarchy problem~\cite{Chacko:2005pe}.
The HV models considered here include a dark sector that extends the SM with a non-Abelian gauge group, similar to quantum chromodynamics (QCD), with new matter and gauge fields analogous to the SM quark and gluon fields.
The SM particles are neutral under the new, dark interaction---``dark QCD''---and communicate with this sector only through a mediator particle that couples to both sectors.

In this Letter, we present the first dedicated search for a characteristic final state consisting of many low-momentum, isotropically distributed particles, which can be described as a soft unclustered energy pattern (SUEP).
This search uses data corresponding to a total integrated luminosity of 138\fbinv from proton-proton collisions at a center-of-mass energy of 13\TeV, collected with the CMS detector at the CERN LHC in 2016--2018.
The concept of SUEPs was first introduced in Ref.~\cite{Strassler:2008bv} and later expanded in Ref.~\cite{Knapen:2016hky} as a signature of HV models.
They had been expected to evade conventional collider data collection strategies and to be difficult to distinguish from energy deposits from multiple proton-proton interactions (pileup).
Similar signatures can arise in other models, such as many-step decays in a hidden sector~\cite{Elor:2015tva}, theories with extra spatial dimensions~\cite{CMS:2010oej,Park:2012fe,Costantino:2020msc}, and instantons~\cite{Amoroso:2020zrz}.

The model that motivates this search includes a quasi-conformal dark sector where the dark QCD force has a large 't Hooft coupling $\lambda \gg 1$ above its confinement scale.
When dark partons are produced, they shower and form dark mesons, whose mass scale is much smaller than the mediator mass scale.
The showering process includes efficient branching over a larger energy range than SM QCD, and all information about the momenta of the initial partons is lost.
The result is a high multiplicity of light HV bound states emitted with a spherical, rather than conical, distribution in the center-of-mass frame of the shower.

\begin{figure}[!htbp]
  \centering
    \includegraphics[width=\figWidth]{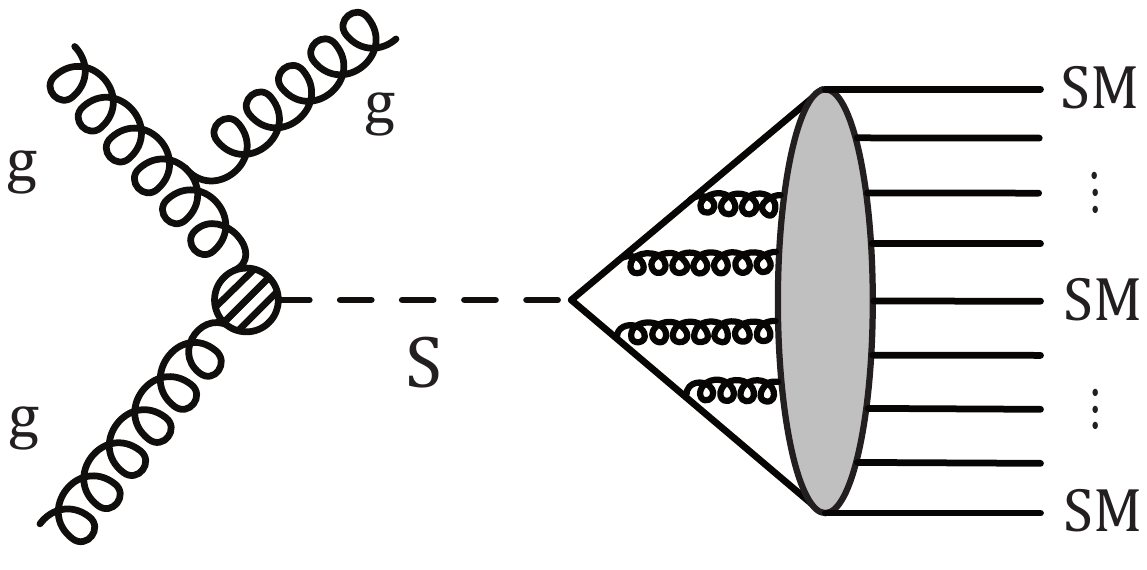}
    \caption{A schematic Feynman diagram of the benchmark signal model resulting in a SUEP signature.}
    \label{fig:feynmanDiagram_ggF}
\end{figure}

This search focuses on a benchmark scenario in which a dark QCD sector is accessed by a heavy scalar mediator \PS, which is produced via gluon fusion and couples directly to a dark quark-antiquark pair.
If the dark quark masses \mqD are below the confinement scale \lambdaD, and $\lambdaD~\ll~\sqrt{s}$, the dark quarks undergo a quasi-conformal showering and form an isotropic spray of dark pseudoscalar mesons \Pphi.
In this strongly coupled model, the \Pphi \pt spectrum can be described by a Boltzmann distribution determined by the mass of the dark meson, \mPhi, and a temperature \TD at the scale of the cutoff, \lambdaD.
Each \Pphi subsequently decays to a pair of dark photons \PAprime.
The dark photon kinetically mixes with the SM hypercharge gauge field and decays promptly to SM particles.
A prompt decay is chosen rather than a displaced decay, which would result in an emerging jet signature~\cite{Schwaller:2015gea,CMS:2024emj}, or extremely long lifetimes, which would result in missing transverse momentum or semivisible jets~\cite{Cohen:2015toa,CMS:2021dzg}.
To enforce promptness, defined as a decay length less than 1\mm, and to avoid dark photon mass (\mAprime) ranges that would already be excluded by collider searches~\cite{Knapen:2016hky}, we choose \mAprime values at the \GeV scale.
The decay branching fraction (\BR) is dictated by the \mAprime value~\cite{Graham:2021ggy}.
We consider three cases with different branching fractions: $\mAprime = 0.5\GeV$ ($\PAprime \to \EE, \MM, \pipi$ with $\BR = 40\%, 40\%, 20\%$), $\mAprime = 0.7\GeV$ ($\PAprime \to \EE, \MM, \pipi$ with $\BR = 15\%, 15\%, 70\%$), and $\mAprime = 1.0\GeV$ ($\PAprime \to \pipi$ with $\BR = 100\%$).
The parameters that describe this dark sector model---\mS, \mPhi, \TD, and \mAprime---can be varied to access a wide spectrum of novel topologies that are characterized by the number, momenta, and particle type of the final-state charged tracks.
A schematic diagram for this signal model is shown in Fig.~\ref{fig:feynmanDiagram_ggF}.

Signal samples are generated using a \PYTHIA plugin based on Refs.~\cite{Knapen:2016hky,Knapen:2021eip}.
Multijet QCD samples used for closure tests are produced with \MGvATNLO 2.6.5~\cite{Alwall:2014hca}.
The \PYTHIA 8.240~\cite{Sjostrand:2014zea} package is used for parton showering, hadronization, and the underlying event simulation, with the event tune CP5~\cite{CMS:2019csb}.
The NNPDF 3.1~\cite{Ball:2017nwa} set of parton distribution functions is used.
The \GEANTfour package~\cite{Agostinelli:2002hh} is employed to perform a detailed simulation of the CMS detector response.
Signal samples are generated with various model parameter values: \mPhi, \TD, \mAprime, and the mediator mass \mS, limited by what is feasibly produced at the LHC; \mAprime is varied as described above.
Low values of $\TD/\mPhi~(\le 0.25)$ are not expected to be realized by any pseudo-conformal hidden sector, while high values constitute the ``dark pion regime,'' where an approximate symmetry generates pseudo-Goldstone bosons~\cite{Barron:2021btf}.

The CMS apparatus~\cite{CMS:2008xjf} is a multipurpose, nearly hermetic detector, designed to trigger on~\cite{CMS:2020cmk,CMS:2016ngn} and identify electrons, muons, photons, and hadrons~\cite{CMS:2020uim,CMS:2018rym,CMS:2014pgm}.
A global ``particle-flow'' algorithm~\cite{CMS:2017yfk} aims to reconstruct all individual particles in an event, combining information provided by the all-silicon inner tracker and by the crystal electromagnetic and brass-scintillator hadron calorimeters, operating inside a 3.8\unit{T} superconducting solenoid, with data from the gas-ionization muon detectors embedded in the flux-return yoke outside the solenoid.
The primary vertex (PV) is taken to be the vertex corresponding to the hardest scattering in the event, evaluated using tracking information alone, as described in Section 9.4.1 of Ref.~\cite{CMS-TDR-15-02}.
The reconstructed particles (particle-flow candidates) are clustered into jets using the anti-\kt algorithm~\cite{Cacciari:2008gp,Cacciari:2011ma} with a distance parameter of $R = 0.4$, after rejecting charged hadrons that do not originate from the PV but rather from pileup interactions; subsequently, neutral pileup contributions are subtracted~\cite{CMS:2016lmd}.

This search uses events that pass a hadronic trigger with a threshold of $\HT > 900~(1050) \GeV$ for 2016 (2017 and 2018), where \HT is defined as the scalar \pt sum of jets with $\pt > 30 \GeV$ and $\abseta < 2.5$.
In addition, an offline selection of $\HT > 1200 \GeV$ is applied to all data to ensure the trigger is efficient.
This implies that all potential SUEP events will have large amounts of hadronic activity.
As a consequence, the SUEP system will recoil against an initial state radiation (ISR) jet with large \pt and therefore will be highly boosted and central in the detector.
In addition to this selection, events are rejected if they contain an electron or muon satisfying $\pt > 25 \GeV$, $\abseta < 2.4$, and isolation requirements~\cite{CMS:2018rym,CMS:2020uim}, in order to avoid any overlap with associated production channels where a \PW or \PZ boson is produced and decays leptonically.

For events passing the trigger, reconstructed charged-particle tracks fitting to the PV (and are therefore not associated with pileup vertices) with $\pt > 0.7 \GeV$ and $\abseta < 2.5$ are clustered into wide jets using the anti-\kt algorithm with a distance parameter of $R = 1.5$.
These wide jets have no additional requirements except to have $\pt > 150 \GeV$, which removes some events associated with soft QCD while retaining a signal efficiency of ${\approx}100\%$.
The two wide jets with the highest \pt are defined as the SUEP-ISR system, where the jet with the larger number of tracks is defined as the SUEP candidate, while the other is defined as the ISR candidate.
The number of constituent tracks in the SUEP candidate jet is defined as \nSUEPConstituents.
These tracks are then boosted into the rest frame of the SUEP candidate, where the scalar mediator decay pattern is expected to be spherically symmetric.
The sphericity of the boosted SUEP candidate is defined using the second and third largest eigenvalues (\lamtwo, \lamthr) of the infrared and collinearly safe generalized sphericity tensor: $\boostedSphericity \equiv \frac{3}{2} \left( \lamtwo + \lamthr \right)$~\cite{Cesarotti:2020ngq}.
The \boostedSphericity can range from 0, maximally non-spherical, to 1, perfectly spherical.
Events where the SUEP candidates have $\nSUEPConstituents < 30$ or $\boostedSphericity < 0.3$ are rejected, while the signal region (SR) is defined to include events with $\nSUEPConstituents > 70$ and $\boostedSphericity > 0.5$.

\begin{figure*}[!htbp]
  \centering
    \includegraphics[width=0.98\linewidth]{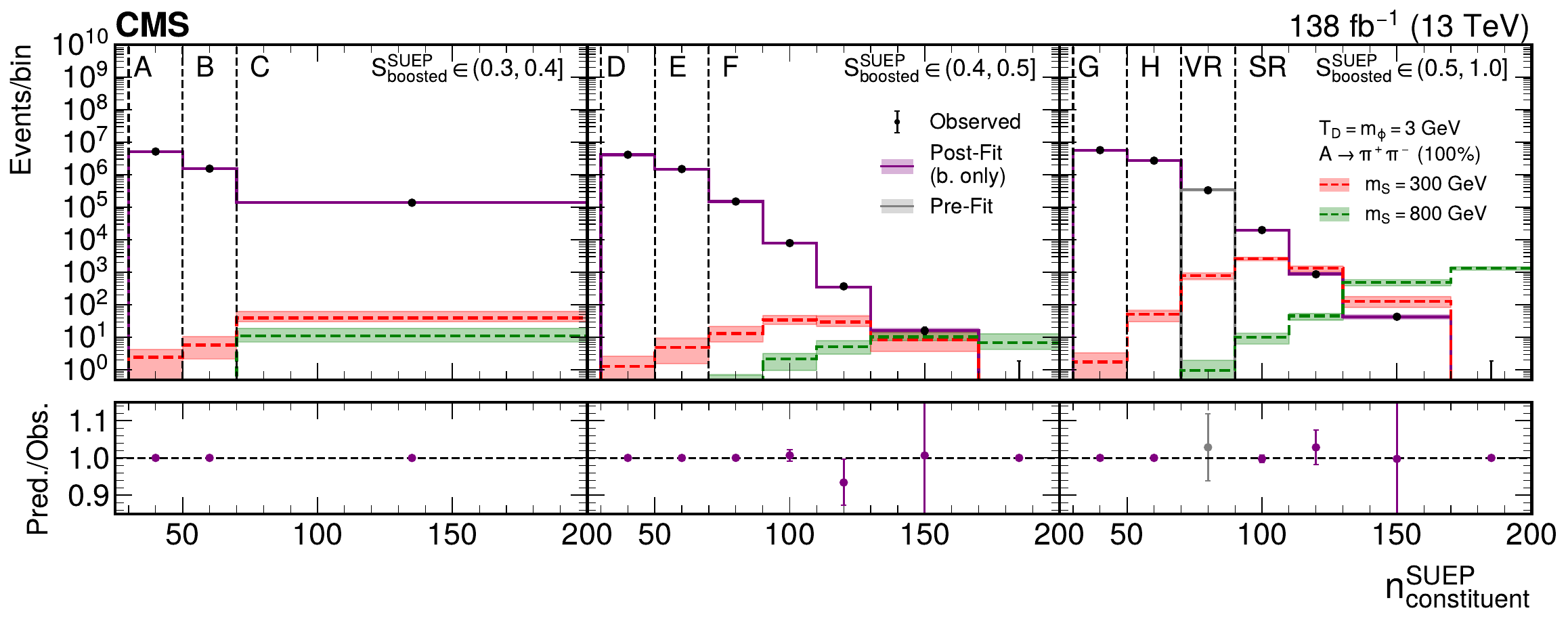}
  \caption{The number of observed events as a function of the number of tracks in the SUEP candidate, for all CRs A--H and the SR, as well as two signal samples. The three figures correspond to contiguous \boostedSphericity ranges. The pre-fit predicted background distribution is shown in the VR, the first bin of the SR. For all other regions and bins, the post-fit values for a background-only fit are shown. \label{fig:all_regions}
  }
\end{figure*}

The primary background consists of events from the QCD multijet process with a large number of tracks.
The background in this search is estimated from data, using the extended ``ABCD'' method proposed in Ref.~\cite{Choi:2019mip}.
The ABCD method makes use of two variables, chosen here to be \boostedSphericity and \nSUEPConstituents.
The classical ABCD method defines one signal region (SR) that includes events passing selection requirements on the two variables and three control regions (CRs) that include events failing one or both selection requirements (``pass-fail'', ``fail-pass'', and ``fail-fail'').
In contrast, to correct for linear correlations between the discriminating variables, the extended ABCD method splits each of the pass-fail and fail-pass CRs into two regions and the fail-fail CR into four regions, resulting in a total of eight CRs, as shown in Fig.~\ref{fig:all_regions}.
In this analysis, the SR is further split into bins of \nSUEPConstituents with edges: 70, 90, 100, 130, 170, $\infty$.
This binning is optimized for sensitivity across the full range of \mS values.
The prediction for each bin $i$ in the SR is computed from the yield $N$ in each of the CRs,
\begin{equation}
  \NSRi = \NFi \frac{ \NF \NH^2 \ND^2 \NB^2 }{ \NGg \NC \NAa \NE^4 }. \label{eq:abcd}
\end{equation}
Because of the large number of QCD background events, the statistical uncertainty is small, $<0.3\%$ in all CRs.
The contamination of SUEP signal events in the CRs is negligible for the same reason.
The accuracy of the background estimation method is verified with simulated QCD multijet samples by testing the agreement between the prediction from simulated CRs and the simulated yield in the SR.
Further, we define the first bin of the SR, which has negligible contribution from signal, as a validation region (VR) in which we perform a closure test.

Systematic uncertainties in the SUEP signal simulation originate from possible mismodeling in the event generation or detector response, and from correction factors that are included when determining the SR selection efficiency.
These include uncertainties in the integrated luminosity measurement for each year of data-taking~\cite{CMS:2021xjt,CMS-PAS-LUM-17-004,CMS-PAS-LUM-18-002} and the reweighting to match the observed pileup distribution.
The variables used in the ABCD method are found not to have substantial dependence on the number of pileup vertices, so an additional uncertainty is not assigned.
There is an uncertainty in the trigger efficiency as a function of \HT; additionally, the uncertainties in the jet energy corrections and jet energy resolution affect the total \HT used to select events.
Uncertainties in the parton showering simulation and the track reconstruction algorithms are applied.
The effect of the track reconstruction efficiency is estimated by comparing yields after artificially removing tracks based on measured reconstruction inefficiencies~\cite{CMS:2023lpp}.
For the model in which the mediator has the same mass as the Higgs boson, additional \pt reweighting corrections are included~\cite{Becker:2020rjp}.
Finally, corrections for a radiation-induced timing shift in the electromagnetic calorimeter readout, resulting in ``prefiring'' of the trigger system, are applied to 2016 and 2017 simulations~\cite{CMS:2020cmk}.
The uncertainties in the parton shower modeling and the \pt reweighting corrections for the $\mS=125\GeV$ case are the largest, corresponding to ${\sim}10\%$ of the signal yield.
The other aforementioned systematic uncertainties correspond to 0\%--4\% of the signal yield.

Further systematic uncertainties in the prediction of the extended ABCD method are estimated from data.
A first systematic uncertainty covers higher-order correlations between the sphericity and number of constituent tracks in the SUEP candidate that are not taken into account by the extended ABCD method.
This is estimated using the ISR candidate, which is a set of tracks distinct from the SUEP candidate in each event.
The ISR candidate is obtained by the same wide-jet clustering as the SUEP candidate in order to test the background estimation method by applying the same selections and ABCD method.
The discrepancy between the total predicted yield and the measured yield in the SR of 8\% is taken as a systematic uncertainty in the normalization value of the predicted background yield.
As shown in Eq.~\eqref{eq:abcd}, the computed prediction in each SR bin $i$ is proportional to $\NFi\NF/\NC$; region $F$ is subdivided, but region $C$ is not, to reduce the statistical uncertainty.
The difference in the \nSUEPConstituents distribution between CRs $F$ and $C$ is used as a proxy for the difference between CR $F$ and the SR, which impacts the final prediction.
The uncertainty is assessed as the ratio $(\NFi/\NF)/(\NCi/\NC)$, measured by binning the $C$ region and using the first two bins, which are the most populated.
The ratio is linearly extrapolated to the less-populated bins as a function of the \nSUEPConstituents centroid for each SR bin and validated in simulation.
This ratio approximates $\NSRi/\NFi$; it is treated as correlated among all bins because of the extrapolation procedure and set to $100\%$ for the last bin, which has no centroid.

\begin{figure}[!htbp]
  \centering
    \includegraphics[width=\figWidth]{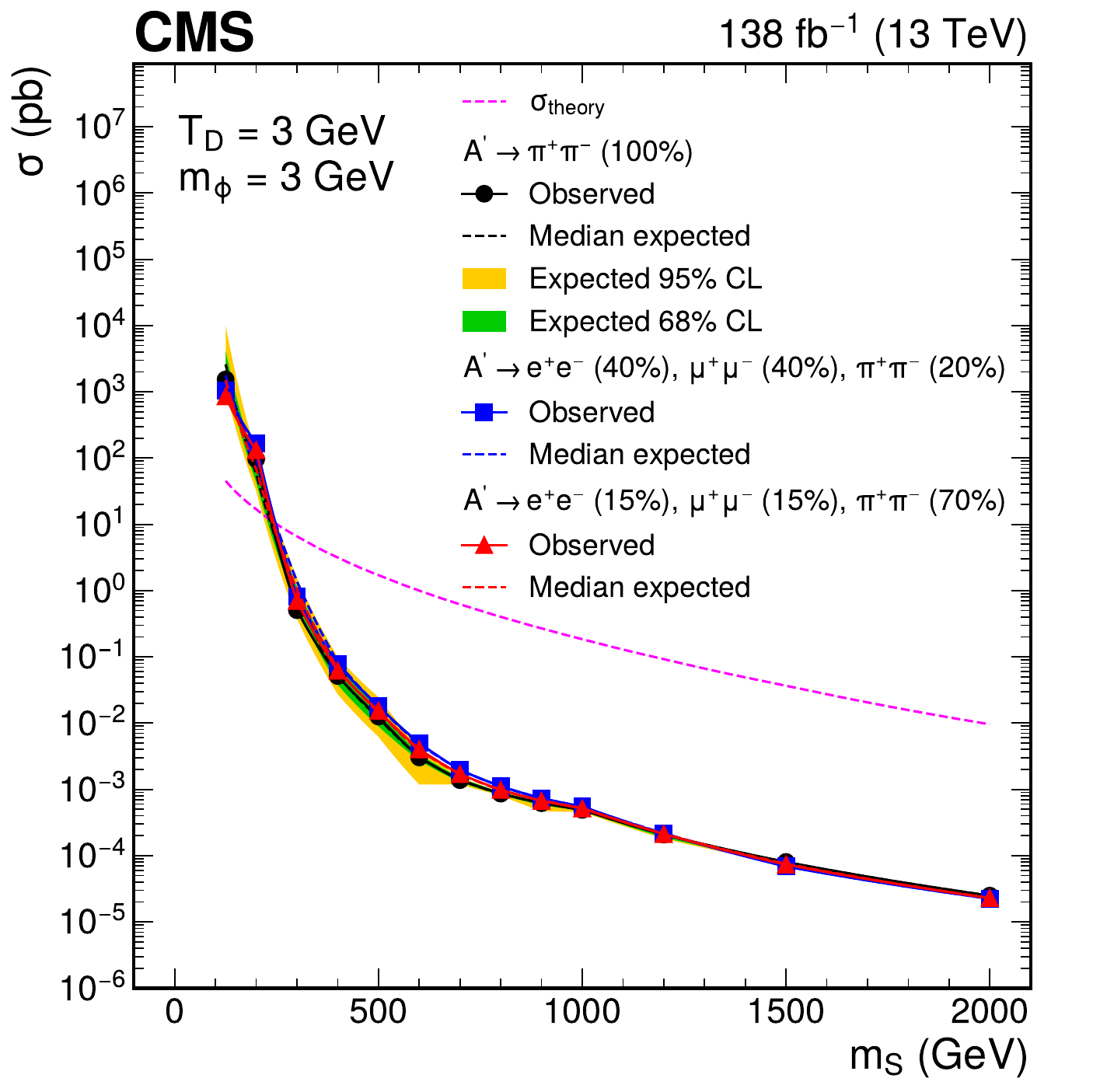}
  \caption{The 95\% \CL exclusion limits on the production cross section $\sigma$ are shown as a function of \mS for $\mPhi = \TD = 3\GeV$, for all decay modes.\label{fig:1D_limits}
  }
\end{figure}

We perform a binned maximum likelihood fit to the \nSUEPConstituents distribution in the SR, excluding the VR.
For each individual SR bin, a Poisson likelihood term is used to describe the statistical fluctuations of the data around the expected central value.
The expected values are calculated by the prediction from the extended ABCD method where the yields in all CRs are allowed to float.
The signal-to-background ratios are different among the three data-taking periods, so they are kept separate in the fit.
The observed and post-fit yields are shown as a function of \nSUEPConstituents in Fig.~\ref{fig:all_regions}.

\begin{figure*}[!hbtp]
  \centering
    \includegraphics[width=0.8\textwidth]{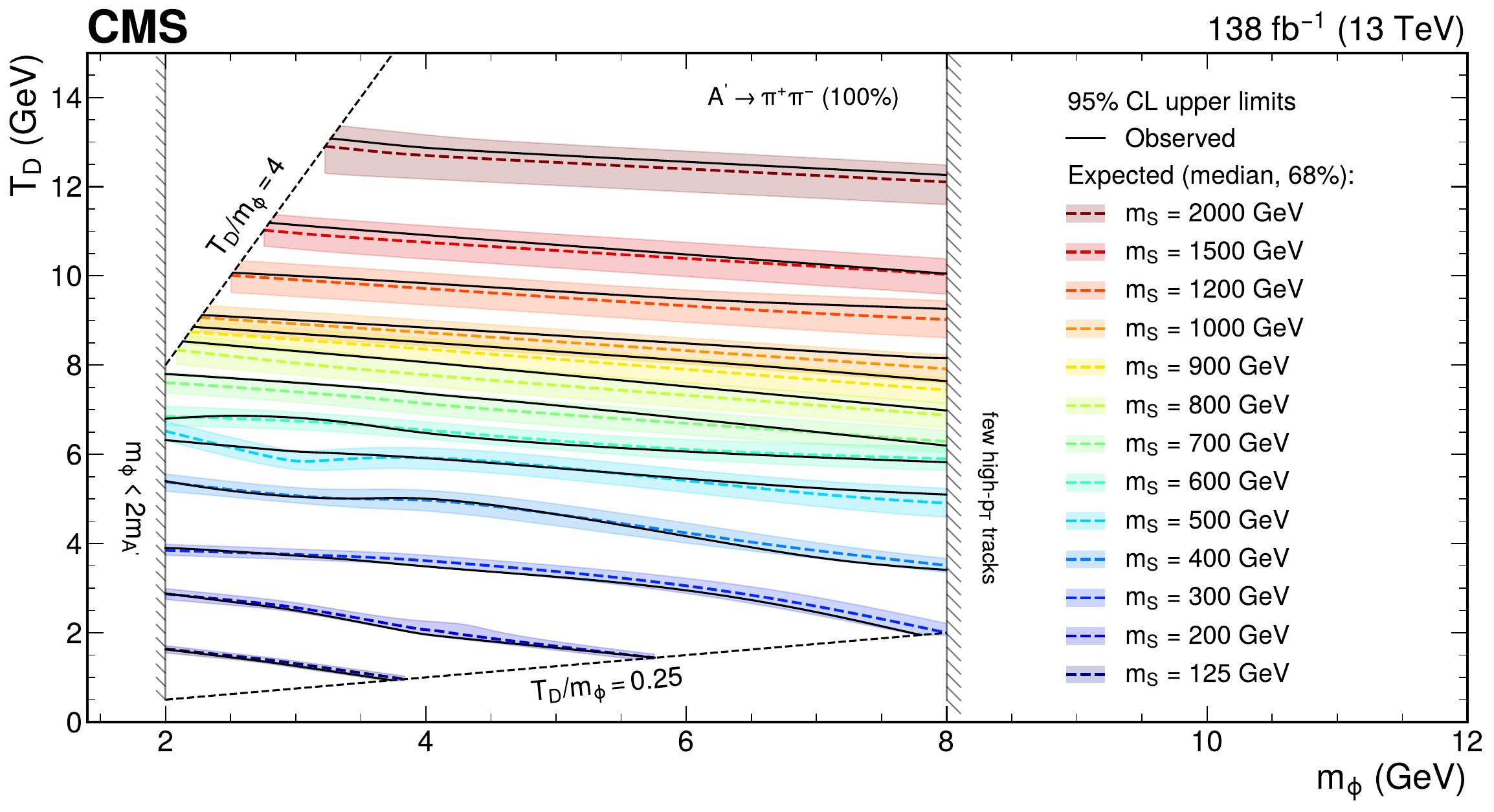}
  \caption{The observed and expected exclusions for the nominal \PS cross section in the plane of \mPhi and \TD, for various \mS values, for the case $\mAprime=1.0\GeV$ ($\PAprime \to \pipi$ with $\BR=100\%$). The regions below the observed limits are excluded.\label{fig:3D_limits}
  }
\end{figure*}

We observe no significant excess of events over the SM prediction.
The post-fit predictions for the $F$ and SR bins with the highest \nSUEPConstituents range are $3.5\times10^{-6} \pm 8.4\times10^{-4}$ and $1.2\times10^{-5} \pm 2.9\times10^{-3}$, respectively.
The modified frequentist construction \CLs~\cite{Junk:1999kv, Read:2002hq} is used to compute 95\% confidence level (\CL) upper limits on the signal cross section.
By comparing these limits to the theoretical cross section for the production of a BSM Higgs boson via gluon fusion through a point-like effective interaction~\cite{deFlorian:2016spz}, ranges of the \mS--\mPhi--\mAprime--\TD parameter space can be excluded.
Figure~\ref{fig:1D_limits} shows the 95\% \CL exclusion limits as a function of \mS for benchmark values of \TD and \mPhi.
There is similar sensitivity to all \PAprime decay cases considered.
Figure~\ref{fig:3D_limits} shows the exclusion limits for various scalar masses in the plane of \mPhi versus \TD, for the case $\PAprime \to \pipi$.
We set stringent limits on the production cross section for the most SUEP-like region of the parameter space with $\mS/\TD \sim 100$ and a final state characterized by a large multiplicity of tracks, which tends to populate the two rightmost bins of the SR, shown in Fig.~\ref{fig:all_regions}.

In summary, this Letter presents the first search for SUEPs.
Data corresponding to an integrated luminosity of 138\fbinv are used, collected with a trigger requiring a high scalar sum of jet transverse momenta and reconstructed with the full offline processing.
This strategy preferentially selects events with initial-state radiation; the characteristic isotropic event shape of the SUEPs is recovered by boosting into the scalar mediator rest frame and removing the initial-state radiation particles.
The number of tracks and the associated sphericity in the SUEP candidate are used to discriminate between the signal and the background from standard model quantum chromodynamics, which is estimated from data in suitable control regions.
Stringent limits are placed on the most SUEP-like Hidden Valley scenarios with highly isotropic dark showers producing a large multiplicity of tracks.

\begin{acknowledgments}
  We congratulate our colleagues in the CERN accelerator departments for the excellent performance of the LHC and thank the technical and administrative staffs at CERN and at other CMS institutes for their contributions to the success of the CMS effort. In addition, we gratefully acknowledge the computing centers and personnel of the Worldwide LHC Computing Grid and other centers for delivering so effectively the computing infrastructure essential to our analyses. Finally, we acknowledge the enduring support for the construction and operation of the LHC, the CMS detector, and the supporting computing infrastructure provided by the following funding agencies: SC (Armenia), BMBWF and FWF (Austria); FNRS and FWO (Belgium); CNPq, CAPES, FAPERJ, FAPERGS, and FAPESP (Brazil); MES and BNSF (Bulgaria); CERN; CAS, MoST, and NSFC (China); MINCIENCIAS (Colombia); MSES and CSF (Croatia); RIF (Cyprus); SENESCYT (Ecuador); ERC PRG, RVTT3 and MoER TK202 (Estonia); Academy of Finland, MEC, and HIP (Finland); CEA and CNRS/IN2P3 (France); SRNSF (Georgia); BMBF, DFG, and HGF (Germany); GSRI (Greece); NKFIH (Hungary); DAE and DST (India); IPM (Iran); SFI (Ireland); INFN (Italy); MSIP and NRF (Republic of Korea); MES (Latvia); LMTLT (Lithuania); MOE and UM (Malaysia); BUAP, CINVESTAV, CONACYT, LNS, SEP, and UASLP-FAI (Mexico); MOS (Montenegro); MBIE (New Zealand); PAEC (Pakistan); MES and NSC (Poland); FCT (Portugal); MESTD (Serbia); MCIN/AEI and PCTI (Spain); MOSTR (Sri Lanka); Swiss Funding Agencies (Switzerland); MST (Taipei); MHESI and NSTDA (Thailand); TUBITAK and TENMAK (Turkey); NASU (Ukraine); STFC (United Kingdom); DOE and NSF (USA).  
\end{acknowledgments}

\textit{Data availability} -- The numerical results from this Letter can be found in its HEPData record~\cite{HEPData}.

\bibliography{auto_generated}
\cleardoublepage \appendix\section{The CMS Collaboration \label{app:collab}}\begin{sloppypar}\hyphenpenalty=5000\widowpenalty=500\clubpenalty=5000
\cmsinstitute{Yerevan Physics Institute, Yerevan, Armenia}
{\tolerance=6000
A.~Hayrapetyan, A.~Tumasyan\cmsAuthorMark{1}\cmsorcid{0009-0000-0684-6742}
\par}
\cmsinstitute{Institut f\"{u}r Hochenergiephysik, Vienna, Austria}
{\tolerance=6000
W.~Adam\cmsorcid{0000-0001-9099-4341}, J.W.~Andrejkovic, T.~Bergauer\cmsorcid{0000-0002-5786-0293}, S.~Chatterjee\cmsorcid{0000-0003-2660-0349}, K.~Damanakis\cmsorcid{0000-0001-5389-2872}, M.~Dragicevic\cmsorcid{0000-0003-1967-6783}, P.S.~Hussain\cmsorcid{0000-0002-4825-5278}, M.~Jeitler\cmsAuthorMark{2}\cmsorcid{0000-0002-5141-9560}, N.~Krammer\cmsorcid{0000-0002-0548-0985}, A.~Li\cmsorcid{0000-0002-4547-116X}, D.~Liko\cmsorcid{0000-0002-3380-473X}, I.~Mikulec\cmsorcid{0000-0003-0385-2746}, J.~Schieck\cmsAuthorMark{2}\cmsorcid{0000-0002-1058-8093}, R.~Sch\"{o}fbeck\cmsorcid{0000-0002-2332-8784}, D.~Schwarz\cmsorcid{0000-0002-3821-7331}, M.~Sonawane\cmsorcid{0000-0003-0510-7010}, S.~Templ\cmsorcid{0000-0003-3137-5692}, W.~Waltenberger\cmsorcid{0000-0002-6215-7228}, C.-E.~Wulz\cmsAuthorMark{2}\cmsorcid{0000-0001-9226-5812}
% ...
\par}
\vskip\cmsinstskip
\dag:~Deceased\\
$^{1}$Also at Yerevan State University, Yerevan, Armenia\\
$^{2}$Also at TU Wien, Vienna, Austria\\
% ...
\end{sloppypar}
%%% END EDITABLE REGION %%%
% skeleton_end
\end{document}